# Quantum Information Processing with Trapped Neutral Atoms

P. S. Jessen[1], I. H. Deutsch[2], and R. Stock[2]

Quantum information can be processed using large ensembles of ultracold and trapped neutral atoms, building naturally on the techniques developed for high-precision spectroscopy and metrology. This article reviews some of the most important protocols for universal quantum logic with trapped neutrals, as well as the history and state-of-the-art of experimental work to implement these in the laboratory. Some general observations are made concerning the different strategies for qubit encoding, transport and interaction, including tradeoffs between decoherence rates and the likelihood of two-qubit gate errors. These tradeoffs must be addressed through further refinements of logic protocols and trapping technologies before one can undertake the design of a general-purpose neutral-atom quantum processor.

**I. Introduction**

An important lessen from twentieth century information science is that "information is physical". One cannot understand the power of algorithms, communication protocols or other information processing tasks separately from the physical description of the devices that perform them. In particular, quantum systems allow the implementation of new types of logic that cannot be efficiently simulated on classical systems governed by laws based on local realism. This has allowed a whole new field to emerge – quantum information science – whose ultimate vision is the construction of a universal quantum computer capable of executing any algorithm that can be described by a quantum evolution.

[1] *Optical Sciences Center, University of Arizona, Tucson, AZ 85721*
[2] *Department of Physics and Astronomy, University of New Mexico, Albuquerque, NM 87131*

Exactly what features give quantum computers their power is still a subject of debate, but certain ingredients are generally agreed upon as essential:

- A many-body system whose Hilbert space has scalable tensor product structure.
- The ability to prepare a fiducial quantum state.
- A universal set of quantum operations capable of implementing an arbitrary quantum map.
- A method to read-out the quantum state.
- A dissipative mechanism to remove the entropy associated with unavoidable errors in a fault-tolerant manner.

Since they were proposed in their original form, we have learned that some of the so-called "DiVincenzo Criteria" [1] can be relaxed. For example, universal quantum maps need not be unitary and may instead have irreversible quantum measurements at their core, as shown by proposals for linear optics quantum computation [2], quantum computation via teleportation [3], and the so-called "one-way quantum computer" in which conditional measurements are performed on an entangled "cluster state" [4]. Such developments highlight an important fact: the roadmap to a universal quantum computer is still evolving, and the "best" way to accomplish a computational task will depend on the strengths and weaknesses of the physical system at hand. Even so, the essential ingredient is clear: quantum control of a many-body system [5], including both reversible unitary evolution and irreversible quantum measurement. Robust, high fidelity execution of these tasks is the goal of all physical implementations of quantum information processing (QIP).

Given these preliminaries, it is clear that atomic, molecular and/or optical (AMO) systems offer unique advantages for QIP. More than in any other subdiscipline, the quantum optics community has explored the foundations of quantum mechanics in the laboratory, including detailed studies of the processes of measurement and decoherence, entanglement and the violation of Bell's inequalities. In appropriately designed dilute systems, coherence times can be very long and decades of research in spectroscopy, precision metrology, laser cooling, and quantum optics has produced a large toolbox with which to manipulate them and drive their quantum dynamics. Indeed, atom- and ion-based atomic clocks are arguably the best controlled, most quantum coherent devices available, and present a strong motivation to consider the use of similar systems for QIP.

## II. Survey

Proposals to use neutral atoms as the building blocks of a quantum computer followed closely after the first demonstration of quantum logic in ion traps [6]. Laser cooling of ions and neutrals was initially developed as an enabling technology for precision metrology. Both systems were known to have long coherence times but also complementary features that lead to radically different approaches to e.g. atomic clock design. Because ions are charged they can be tightly confined in deep traps and observed for very long times, but the strong Coulomb repulsion limits the number of ions that can be precisely controlled in a single trap. In contrast, neutral atoms usually interact only at very short range and can be collected in large ensembles without perturbing each other, a clear advantage for both metrology and QIP. On the downside, traps for neutrals are shallow compared to ion traps, and the atom/trap field interaction invariably perturbs the atomic internal state. In QIP one must balance an intrinsic conflict – qubits must interact with each other and with external control fields that drive the quantum algorithm, while at the same time the system must couple only weakly to the noisy environment which leads to decoherence. In an ion trap the Coulomb interaction leads to collective modes of center-of-mass motion, which can be used as a "bus" for coupling qubits together [6]. However, control of a strongly coupled many-body system becomes increasingly complex as the system size grows, and will likely require the use of intricate multi-trap designs to overcome the difficulty of working with even a handful of ions in a single trap [7]. Also, the strong interactions can have a parasitic effect by coupling the ionic motion to noisy electric fields such as those associated with patch potentials on the trap electrodes [8]. Neutral atoms in the electronic ground state, in contrast, couple weakly to each other and to the environment, and so offer a different compromise between coupling vs. control complexity and decoherence.

The generally weak- and short-range coupling between neutrals makes the introduction of non-separable two-qubit interactions *the critical element* of neutral atom QIP. Brennen *et al.* [9] and Jaksch *et al.* [10] realized independently that this might be achieved by encoding qubits in the hyperfine ground manifold of individual atoms trapped in optical lattices [11], and using the state sensitive nature of the trap potential to bring the atomic center-of-mass wavepackets together for controlled interactions mediated by either optical dipole-dipole coupling [9] or ground state collisions [10]. Further ideas include a proposal for fast quantum gates based on interactions between Rydberg atoms [12], and another based on magnetic spin-spin interaction [13]. These developments occurred against a backdrop of steady progress in the technologies for cooling, trapping and manipulating neutrals, in particular in optical lattices. Early work that helped inspire proposals for QIP include the demonstration of Raman sideband cooling to the lattice vibrational ground state [14], the generation of vibrational Fock- and delocalized Bloch-states [15], and tomographic reconstruction of the atomic internal [16] and center-of-mass state [17]. At the same time theoretical work indicated that loading an optical lattice from a Bose-Einstein condensate can induce a transition to a Mott-insulator state with nearly perfect, uniform occupation of the lattice sites [18]. A series of ground-breaking experiments by the group of Bloch and Hänsch have recently demonstrated, in short order, first the Mott-insulator transition [19], followed by coherent splitting and transport of atomic wavepackets [20], and finally controlled ground-ground state collisions and the generation of entanglement in an ensemble consisting of short strings of atoms [21]. Other elements of neutral atom QIP have been pursued in a number of laboratories, including patterned loading of optical lattices [22], addressing of individual lattice sites [23], and alternative trap technologies such as magnetic microtraps [24] and arrays of optical tweezers traps [25,26].

**II.A. Neutral atom traps**

Implementation of neutral atom QIP is closely tied to the development of suitable traps. Neutral atom traps in general rely on the interaction of electric or magnetic dipole moments with AC and/or DC electromagnetic fields. Magnetic traps have found wide use in the formation of quantum degenerate gases, but tend to be less flexible than optical traps in terms of the atomic states that can be trapped, and therefore have not been as widely considered for QIP. For this reason we concentrate on optical traps created by the dynamical (AC) Stark effect in far detuned, intense laser fields. In principle these traps suffer from decoherence caused by the spontaneous scattering of trap photons, but in practice the rate can be suppressed to a nearly arbitrary degree through the use of intense trap light tuned very far from atomic resonance. Proposals for QIP typically have considered alkalis (e. g. Rb or Cs) which are easy to laser-cool and have nuclear spin so qubits can be encoded in long-lived hyperfine ground states. For these atomic species trap detunings are always much larger than the excited state hyperfine splitting. In this limit the optical potential can be written in the compact form [27], $U(\mathbf{x}) = U_s(\mathbf{x}) - \boldsymbol{\mu} \cdot \mathbf{B}_{fict}(\mathbf{x})$, where $U_s(\mathbf{x})$ is a scalar potential (independent of the atomic spin) proportional to the total laser intensity, and $\mathbf{B}_{fict}$ is a *fictitious* magnetic field that depends on the *polarization* of the trap light, and $\boldsymbol{\mu} = g_F \mu_B \mathbf{F}$, where $\mathbf{F}$ is the total angular momentum (electron plus nuclear) and $g_F$ is the Landé $g$-factor. For trap detunings much larger than the excited state fine structure $\mathbf{B}_{fict} \to 0$, and the potential is always purely scalar.

This description is the foundation for designing QIP protocols. To illustrate this point we consider how to bring atoms together for controlled interactions in a one-dimensional (1D) optical lattice consisting of a pair of counterpropagating plane waves whose linear polarizations form an angle $\theta$ (Fig. 1). Choosing the $z$-axis along the lattice beams, the optical potential is given by $U_s(\mathbf{x}) = 2U_0(1 + \cos\theta \cos 2kz)$, $\mu_B \mathbf{B}_{fict} = U_0 \sin\theta \sin 2kz \, \mathbf{e}_z$, where $U_0$ is the light shift in a single, linearly polarized lattice beam and $k$ the laser wave number. For $\sin(\theta) \neq 0$ there is a gradient of the fictitious **B**-field near the minima of the scalar potential $U_s(\mathbf{x})$, which separates the different magnetic sublevels as in a Stern-Gerlach apparatus and causes the trap minima for hyperfine substates $|F, \pm m_F\rangle$ to move in opposite directions along $z$. A closer inspection of the full lattice potential shows that the trap minima move by $\pm \lambda/2$ for every $2\pi$ increase of the polarization angle $\theta$. Thus, a pair of atoms in e. g. $|F, m_F\rangle$ and $|F, -m_F\rangle$, trapped in neighboring wells at $\theta = \pi/2$, can be superimposed by rotating the lattice polarization to $\theta = \pi$, and separated again by further polarization rotation.

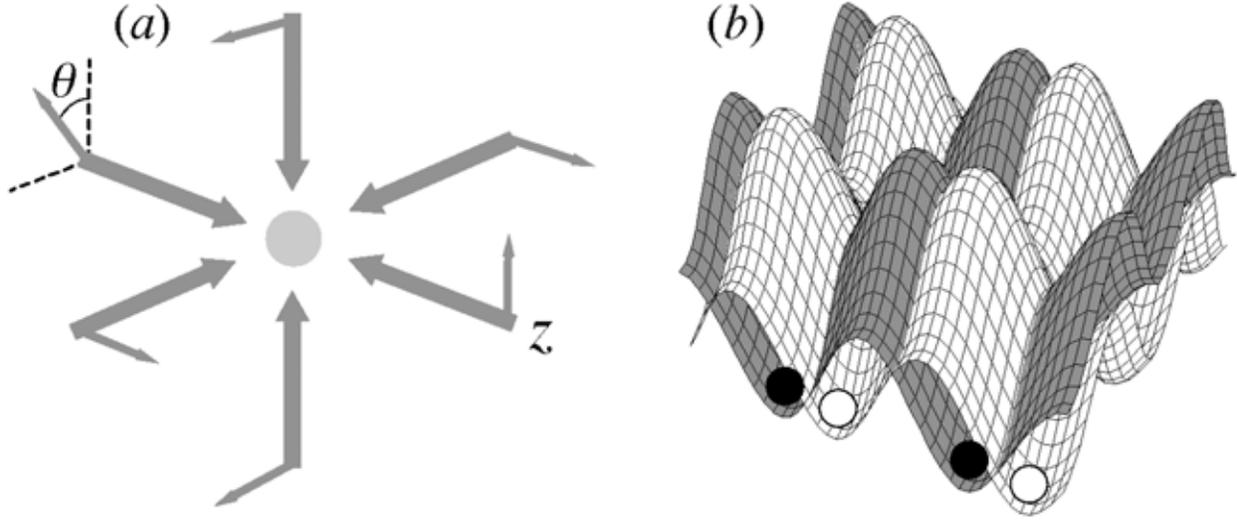

Fig. 1: Schematic of a 3D optical lattice. (a) Two pairs of linearly polarized beams provide transverse confinement, and the beams along $z$ in the lin-$\theta$-lin configuration provide longitudinal confinement in $\sigma_+$ and $\sigma_-$ standing waves. (b) Potential surfaces for the atom in different magnetic sublevels, described in the text, shown here as in gray and white, are moved along the $z$-axis through a rotation of the angle $\theta$ between polarization vectors for controlled collisions.

**II.B. Quantum logic**

The basic design of a QIP protocol in the standard quantum circuit model involves a choice of qubit encoding, initialization method, single- and two-qubit gates, and read-out method. Of these mutually dependent design elements, the implementation of unitary two-qubit entangling gates poses the most fundamental challenge. One well known example of a universal two-qubit gate is the controlled-phase (CPhase) gate, which maps the two-qubit logical basis state $|1\rangle|1\rangle \to -|1\rangle|1\rangle$, and leaves the others unchanged. In fact, any gate based on a diagonal two-qubit Hamiltonian can be converted to CPhase by single-qubit rotations, provided that the energy shifts are non-separable, $\Delta E \equiv E_{11} + E_{00} - (E_{10} + E_{01}) \neq 0$, and the duration of the interaction is $\tau = \pm \pi \hbar/\Delta E$. If noise and/or decoherence introduces errors at a rate $\gamma$ then we can estimate the minimum error probability of such a gate, $P_{error} = 1 - e^{-\gamma\tau} \approx \pi \hbar \gamma/\Delta E$. The quantity $\Delta E/\gamma$ is thus a key figure of merit of the gate operation, with a clear physical interpretation; it is the *spectral resolvability* of the coupled two-qubit states.

Because of their short range, neutral-atom interactions are best understood in terms of controlled collisions. To implement high-fidelity quantum logic these collisions must be state dependent, but at the same time they must not cause scattering into states outside the computational basis. In atomic systems these requirements are generally in conflict, but can be reconciled through appropriate choices of qubit encoding and trap geometry. Jaksch *et al.* proposed to use elastic *s*-wave collisions of atoms in the electronic ground state [10]. In this protocol the main concern is to suppress inelastic collisions caused by the Heisenberg spin-exchange interaction that preserves only the *total* magnetic quantum number, but not that of the individual atoms. Jaksch *et al.* solved this problem by encoding qubits in the stretched states $|1\rangle = |F_+, m_F = F_+\rangle$, $|0\rangle = |F_-, m'_F = F_-\rangle$, where $F_\pm = I \pm 1/2$. Because $g_{F_\pm} = \pm 1/F$ these states move in opposite directions in a lattice of the type discussed in II.A. Rotating the lattice polarization angle from $\theta = 0$ to $\pi$ will then cause at atom in the state $|0\rangle$ and moving to the right to collide with an atom in the state $|1\rangle$ and moving to the left, i. e. the two qubits interact only if the state is $|0\rangle|1\rangle$ and not otherwise. In that case $\Delta E = E_{01} \neq 0$ and a CPhase can be achieved. Furthermore, because *s*-wave scattering conserves $m_F + m'_F$ (to good approximation) and neither $m_F$ nor $m'_F$ can increase, this collision *must* be elastic.

Several additional protocols for two-qubit interactions have been proposed. For example, Charron *et al.* [28] and Eckert *et al.* [29] considered encoding qubits in the ground and first excited center-of-mass vibrational states of trapped atoms, and to couple atomic qubits in neighboring traps by lowering the intervening potential barrier until tunneling causes atoms in the excited states to couple via *s*-wave collisions. Brennen *et al.* considered collisions of nearby but non-overlapping wavepackets associated with different internal states in different potentials [9]. This gives greater flexibility to design elastic but state-dependent interactions, but requires resonant and/or longer-range forces than the $1/r^6$ van der Waals potential between ground state atoms. Brennen *et al.* proposed to use the $1/r^3$ electric dipole-dipole interactions created when an off-resonant laser field mixes the ground-state manifold with excited electronic states. These excited states will spontaneously emit photons and cause errors, but the rate saturates to that of the two-atom superradiant state when the atoms are separated by less than a wavelength, while the dipole-dipole interaction continues to increase with decreasing atomic separation. Thus, for very tightly localized wavepackets in close proximity, the dipole-dipole interaction can be nearly coherent. Relatively long-range interactions provide yet another strategy to implement quantum logic with neutrals [12]. If atoms are excited into high-lying Rydberg states one can induce very large dipole moments by applying a static electric field. The interaction between two such dipoles is large enough to provide useful level shifts even if atoms are separated by several microns. In one possible protocol, qubits are encoded in the magnetic-field insensitive "clock doublet", $|1\rangle = |F_+, m_F = 0\rangle$, $|0\rangle = |F_-, m_F = 0\rangle$. To execute a two-qubit gate the atoms are excited by a laser tuned to the transition from the logical state $|1\rangle$ to a Rydberg level. If the atoms are not too far separated the Rydberg dipole-dipole interaction is strong enough to shift the two-atom, doubly excited state out of resonance and prevent it from becoming populated, a phenomenon referred to as "dipole-blockade". Since the blockade occurs only for the $|1\rangle|1\rangle$ logical state it can be used to achieve a CPhase.

**II.C Experimental progress.**

Efforts to implement neutral atom QIP in the laboratory represent a natural but challenging extension of existing tools to prepare, control and measure the quantum state of trapped neutrals. A number of experiments have demonstrated several of the key components that go into QIP, and very recently some of these have been combined for the first time to demonstrate control and entanglement in a neutral-atom many body system. In this section we briefly review progress in three main areas: initialization of the qubit register, implementation of single- and two-qubit gates, and methods to address individual qubits.

Optical lattices typically confine atoms tightly on the scale of an optical wavelength (the Lamb-Dicke regime), and lend themselves readily to the use of Raman sideband cooling. In a first demonstration, Hamann *et al*. initialized 98% of a $10^6$-atom ensemble in a single spin- and vibrational-ground state of a sparsely filled 2D lattice [14], and subsequent work has achieved a somewhat lesser degree of state preparation in nearly filled 3D lattices [30]. These laser cooling-based approaches are relatively simple to implement and will work in any tightly confining trap geometry, but when used in a lattice will produce a random pattern of vacant and occupied sites. Sparse, random filling may suffice for ensemble-based investigations of quantum logic [31], but falls short of the requirements of full-scale lattice-based QIP.

Better filling and initialization can be achieved by loading a 3D lattice from a high-density Bose-Einstein condensate and driving the atom/lattice through a superfluid to Mott insulator phase transition [18]. The group of Bloch and Hänsch at MPQ in Münich used this approach as a starting point for a series of proof of principle experiments to establish the viability of the Jaksch *et al.* collisional protocol [10]. As the first step, Greiner *et al*. successfully demonstrated the transition to an "insulator" phase consisting of individual $^{87}$Rb atoms localized in the ground state of separate potential wells [19]. Mandel *et al.* then explored spin-dependent coherent transport in the context of interferometry [20]. This was done by preparing atoms in the logical-$|0\rangle$ state, transferring them to an equal superposition of the states $|0\rangle$ and $|1\rangle$ with a microwave $\pi/2$ pulse, and "splitting" them into two wavepackets by rotating the laser polarization vectors. The "which way information" was then erased with a final $\pi/2$ pulse and the atoms released from the lattice, allowing the separated wavepackets of each atom to overlap and interfere as in a two-slit experiment. Inhomogeneities across the ensemble were at least partially removed through a spin-echo procedure using additional $\pi$ pulses. In this fashion the experiment achieved fringe visibilities of 60% for separations of three lattice sites, limited by quantum phase-errors induced by magnetic field noise, vibrational heating and residual inhomogeneities. Finally, Mandel *et al.* performed a many-body version of this experiment in a nearly filled lattice [21], where the majority of atoms underwent collisional interactions with their neighbors according to the Jaksch *et al.* protocol. For appropriate collision-induced phase shifts this will lead to the formation of chains of entangled atoms, which cannot then be disentangled again by "local" operations such as the final $\pi/2$ pulse. In the experiment a periodic disappearance and reappearance of interferometer fringe visibility was clearly observed as a function of interaction time and corresponding degree of entanglement. Technical limitations, in particular the inability to perform single qubit measurements, have so far made it difficult to obtain quantitative estimates for the size and degree of entanglement of these cluster states, or to extract the fidelity of the underlying CPhase interaction.

The experiments just described are essentially multiparticle interferometry, and illustrate how proof-of-principle and optimization of a gate protocol can be achieved with ensemble measurements. To proceed towards universal QIP it will be necessary to develop an ability to manipulate and read out the state of individual atomic qubits. In principle this can be accomplished by performing single-qubit rotations with focused Raman beams rather than microwave fields, and single-qubit readout with focused excitation beams and/or high-resolution fluorescence imaging. However, the necessary optical resolving power will be nearly impossible to achieve in current lattices whose sites are separated by roughly 0.5 μm. There are several possible ways around this problem: the lattice can be formed by a $CO_2$ laser so individual sites are 5 μm apart and resolvable with a good optical microscope [23], or a conventional lattice can be loaded with a pattern where atoms occupy only every *n*'th well [22]. Alternatively, one might use other trapping geometries, such as arrays of very tightly focused optical tweezers-type traps. Schlosser *et al.* has shown that a few such traps can be formed in the focal plane of a single high-NA lens, and that the trap lens can be used at the same time to achieve spatially resolved detection of fluorescence [25]. This work used the ability to detect single atoms, in combination with a phenomenon known as "collisional blockade", to load individual traps with exactly one atom each. Much larger arrays of such traps have been demonstrated using microfabricated arrays of high-NA microlenses [26], but this approach has yet to demonstrate the loading and detection of one atom per trap.

## III. Lessons Learned and Future Research

The seminal experiments by the Münich group have demonstrated the feasibility of coherent spin transport and entanglement via controlled collisions, but also served to highlight some of the fundamental limitations of the particular protocol employed. To implement high-fidelity collisional gates one must achieve a spin-dependent phase shift, while at the same time restrict the interaction to a single collisional channel so as to prevent scattering outside the computational basis. Jaksch *et al.* accomplished this with their stretched-state encoding, but at the cost of being maximally sensitive to magnetic field- and trap noise which was already a limiting factor in the Münich experiments. Moreover, in a filled lattice the protocol leads to large entangled chains rather than the isolated two-qubit interactions required in the standard quantum circuit model.

It is of course conceivable that one might switch between-noise protected encodings and encodings suitable for collisions during the course of a computation, but such an approach would be cumbersome. Our group is now exploring an alternative, by developing new methods to accurately control collisions between cold atoms in tight traps. As in the original proposal by Brennen *et al.*, we consider logical basis states $|0\rangle = |F_+, m_F\rangle$ and $|1\rangle = |F_-, -m_F\rangle$ for which Zeeman and AC Stark shifts are close to identical. With such encodings the logical states move on identical optical potentials and are never split into separated wavepackets. This provides excellent immunity against noise, but at a cost: in a two-qubit interaction all four logical states interact. The challenge is then to engineer a collision to produce a non-separable phase shift without inelastic scattering. The possibilities of coherent control by directly manipulating the center of mass wave packets for atoms in tight traps offer new avenues to reach this goal. A particularly promising approach is to consider *resonant* interactions between atoms in spatially separated traps that can then be used to pick out and strengthen a single elastic channel and suppress off-resonance inelastic processes.

Stock *et al.* have studied the resonant interaction that occurs when a molecular bound state is AC Stark shifted into resonance with a center-of-mass vibrational state of the two-atom system [32]. These "trap-induced shape resonances" show up as avoided crossings in the energy spectrum as a function of the trap separation, as shown in Fig. 2. The energy gaps indicate the strength of the resonance and become substantial when the scattering length associated with the collision is on the order of the trapped wave packet's width. At this point the two-atom interaction energy is a nonnegligible fraction of the vibrational energy. The Münich experiments used $^{87}$Rb atoms for which the relevant scattering length is ~100 $a_0$, and a shallow lattice potential where the trapped wave packet width was ~1200 $a_0$, resulting in a negligible energy gap of order $10^{-22}$ $\hbar\omega$. If we choose to work instead with $^{133}$Cs, the relevant scattering length lies in the range from 280 $a_0$ to 2400 $a_0$, which is comparable to the ~200 $a_0$ wave packet width in a moderately deep lattice. In this case the trap induced shape resonance will be significant, and should provide a new and flexible mechanism for designing quantum logic protocols. Additional flexibility and control can in principle be introduced by tuning the scattering length via optically or magnetically induced Feshbach resonances, as demonstrated in several BEC experiments [33].

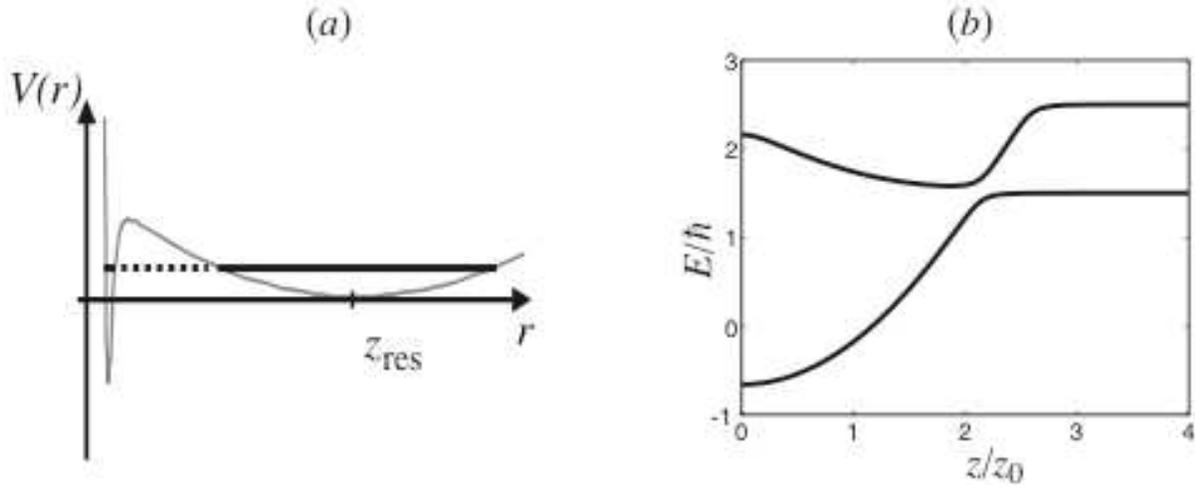

Fig 2. (a) Sum of the harmonic trapping potential and chemical binding potential (gray line), as a function of the relative coordinate $r$ along a line through the two trap minima. The trap eigenstate can become resonant with a molecular bound state at a critical separation $\Delta z_{res}$. (b) The energy spectrum as a function of separation between traps $\Delta z$ (in units of the trap ground state width $z_0$) shows the energy shift of the molecular bound state due to the harmonic trapping potential and the avoided crossings associated with the trap induced resonance.

The Jaksch *et al.* proposal and Münich experiments together provide proof-of-principle that the most important components of QIP can be achieved with trapped neutral atoms, but are still far from a full quantum computer architecture. Spin dependent trapping forces are at the heart of the protocol, and the trap detuning therefore can be at most comparable to the excited state fine structure. The resulting photon scattering ultimately leads to motional heating, decoherence, and even the occasional loss of an atom. It is therefore necessary to explore mechanisms for re-cooling and replacing atoms, and to provide a supply of fresh ancilla atoms as required for error correction. Most importantly, trapping architectures must be developed that allow efficient, programmable transport and qubit interaction, along with individual qubit manipulation and readout. Long-period or pattern loaded [22] lattices or arrays of tweezers traps are one step in this direction, as is recent work on microwave spectroscopy in micro-magnetic traps [34]. Protocols based on Rydberg atoms provide additional freedom to design a workable QIP architecture [12]. Because of the longer range of the interaction there is in principle no need for spin dependent transport, and trap fields can therefore be detuned much further from resonance. This should effectively remove one important source of heating and decoherence. However, the approach raises new challenges related to the coherent control of Rydberg atoms, e. g. accurate and highly coherent $\pi$-pulses between ground and Rydberg levels. Rydberg atoms are also highly susceptible to background DC and AC electric fields, as well as to spontaneous decay and perturbation by thermal blackbody radiation.

As the review and discussion in this article illustrates, both the details and overall architecture of a hypothetical neutral atom quantum processor continues to evolve. Every known approach involves tradeoffs between conflicting requirements, and much additional research is required before we can hope to identify a winning strategy. In addition, new paradigms are being developed, inspired by the physical constraints of the particular implementations under study. An excellent example is the "one-way quantum computer" of Raussendorf andBriegel, in which the type of cluster stats generated in the Münich experiments become a resource for computation rather than a liability [4]. Whether this protocol can be made fault tolerant is a subject of continued research. Indeed, fault tolerance is the ultimate goal of any QIP implementation, and it will eventually be necessary to consider in detail how it might be achieved in the context of concrete logic protocols and architectures. Optical lattices and similar traps that allow blocks of physical qubits to be encoded and manipulated in parallel provide an attractive architecture for error correction. More speculatively, error correction based on topological codes might be implemented in a lattice geometry [35] and lead to a very robust fault-tolerant architecture. Which, if any of these ideas ultimately turn out to be practical remains to be seen. Clearly, information is still physical.

**Acknowledgments:** Research at UA was partly supported by the Defense Advanced Research Projects Agency (DARPA) under Army Research Office (ARO) Contract No. DAAD19-01-1-0589, and by the National Science Foundation under Contract No. ITR-0113538. Research at UNM was partly supported by the National Security Agency (NSA) and the Advanced Research and Development Activity (ARDA) under Army Research Office (ARO) Contract No. DAAD19-01-1-0648 and by the Office of Naval Research under Contract No. N00014-00-1-0575.